\pdfoutput=1
\documentclass{iau}
\usepackage{graphicx}

\title[Non-geodesic orbital frequencies in vicinity of magnetized neutron stars] %% give here short title %%
{Non-geodesic orbital and epicyclic frequencies in vicinity of slowly rotating magnetized neutron stars}

\author[P. Bakala et al.]   %% give here short author list %%
{Pavel Bakala$^1$, Martin Urbanec$^1$, Eva {\v{S}}r{\'{a}}mkov{\'{a}}$^1$,\\
Zden{\v e}k Stuchl\'{\i}k$^1$ and
Gabriel T\"or\"ok$^1$}

\affiliation{$^1$Institute of  Physics, Faculty of Philosophy and Science, Silesian University in Opava \\
    Bezru{\v c}ovo n\'am. 13, CZ-746 01 Opava, Czech Republic  \\ email: {\tt pavel.bakala@fpf.slu.cz } }

\pubyear{2012}
\volume{290}  %% insert here IAU Symposium No.
\jname{Feeding compact objects: Accretion on all scales}
\editors{C.M. Zhang, T. Belloni, M. M\'endez \& S.N. Zhang, eds.}

\begin{document}

\maketitle

\begin{abstract}
We study non-geodesic corrections to the quasicircular motion of charged test particles in the field of magnetized slowly rotating neutron stars. The gravitational field is approximated by the Lense-Thirring geometry, the magnetic field is of the standard dipole character. Using a fully-relativistic approach we determine influence of the electromagnetic interaction (both attractive and repulsive) on the quasicircular motion. We focus on the behaviour of the orbital and epicyclic frequencies of the motion. Components of the four-velocity of the orbiting charged test particles are obtained by numerical solution of equations of motion, the epicyclic frequencies are obtained by using the standard perturbative method. The role of the combined effect of the neutron star magnetic field and its rotation in the character of the orbital and epicyclic frequencies is discussed.
\end{abstract}

%\vspace{-0.5cm}

\begin{figure}[t!]
\centering
\includegraphics[width=1.0\hsize, angle=0]{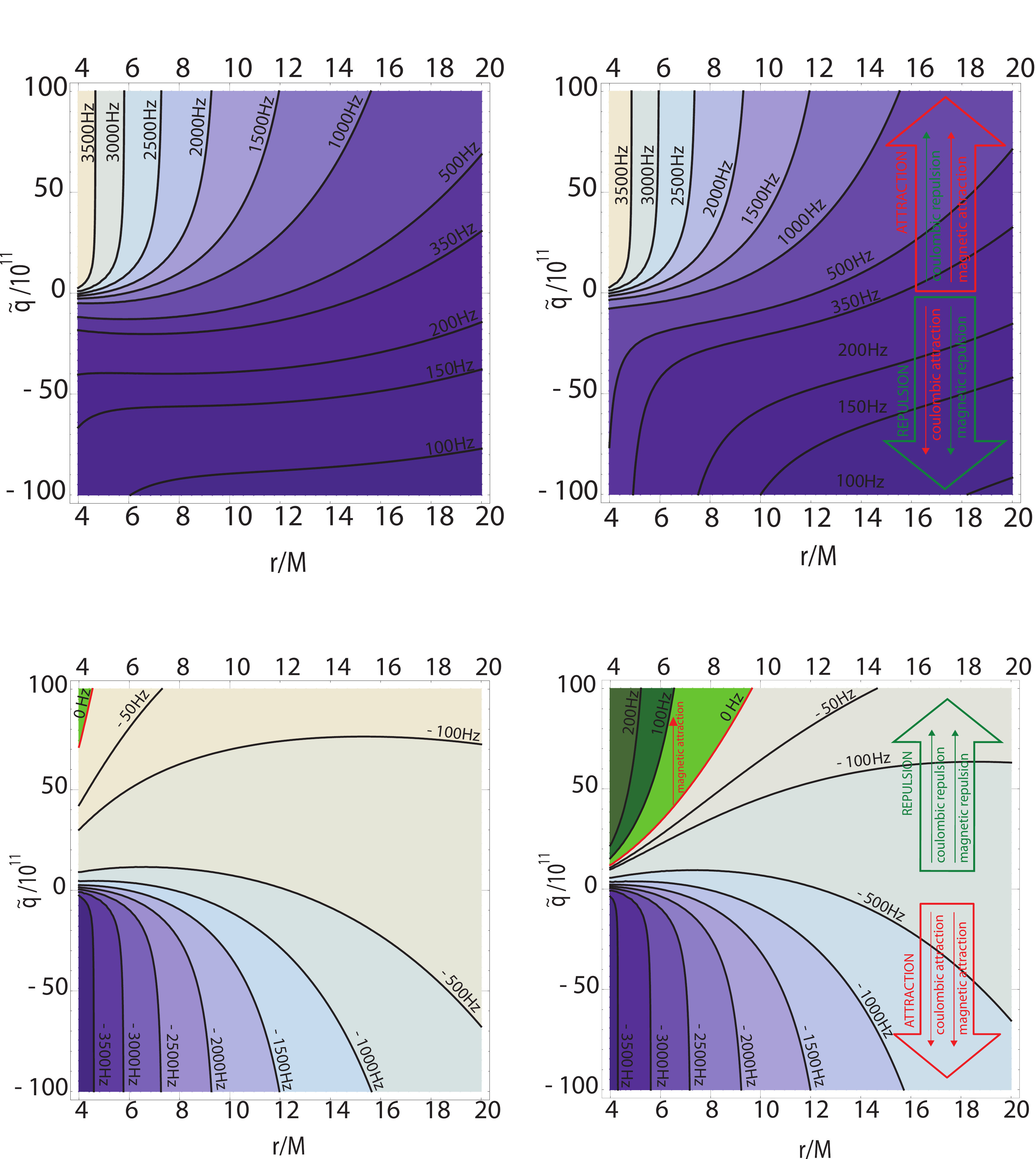}
\caption[t]
{Contour plot of the orbital frequency $\nu=\omega / 2\pi$ as a function of the specific charge $\mathrm{\tilde q}$ and the radial coordinate.
Top left: Corotating solution for spin $a=0.05$. Top right: Corotating solution for spin $a=0.3$.
Bottom left: Counter-rotating solution for spin $a=0.05$. Bottom right: Counter-rotating solution for spin $a=0.3$.}
\label{figure:omega}
\end{figure}

We assume the external gravitational field of slowly rotating neutron or strange stars approximated by the Lense-Thirring metric (\cite[Lense \& Thirring 1918, Hartle \& Sharp 1967, Hartle 1967]{len-thir:1918,hartle-sharp:1967,hartle:1967}), the magnetic dipole with the symmetry axis identical with rotation axis and infinitely conductive star interior implying force lines frozen into the star and dragged by its rotation (\cite[Konno \& Kojima 2000]{kon-koj:2000}). Solving the equation of motion for a charged test particles, we obtain orbital angular velocities $\omega\pm$ for corotating and counter-rotating equatorial circular orbits (Fig. \ref{figure:omega}). Properties of the corotational and counter-rotational circular orbital motion are naturally different due to the frame dragging of the external spacetime. The asymmetry is further amplified by the existence of two components of the Lorentz force with different behaviour. The Coulombic part of the interaction is induced by star rotation only, and is therefore independent of the test particles orbital velocity, while the orientation and amplitude of the magnetic part of the interaction induced by the orbital motion are strongly dependent on the orbital velocity.

Surprisingly, in the case of the counter-rotating orbits, starting from a critical specific charge $\mathrm{\tilde q_{es}}$, the character of the electromagnetic interaction enables the existence of the \hbox{\textbf{electrostatic radius $r_{es}(\mathrm{\tilde q})$}}, where properly charged particles are static relative to static observers at infinity. In such case, the gravitational attraction of the neutron star is compensated by the electric repulsion (Fig. \ref{figure:omega}).

\begin{figure}[b!]
\includegraphics[width=1.0\hsize, angle=0]{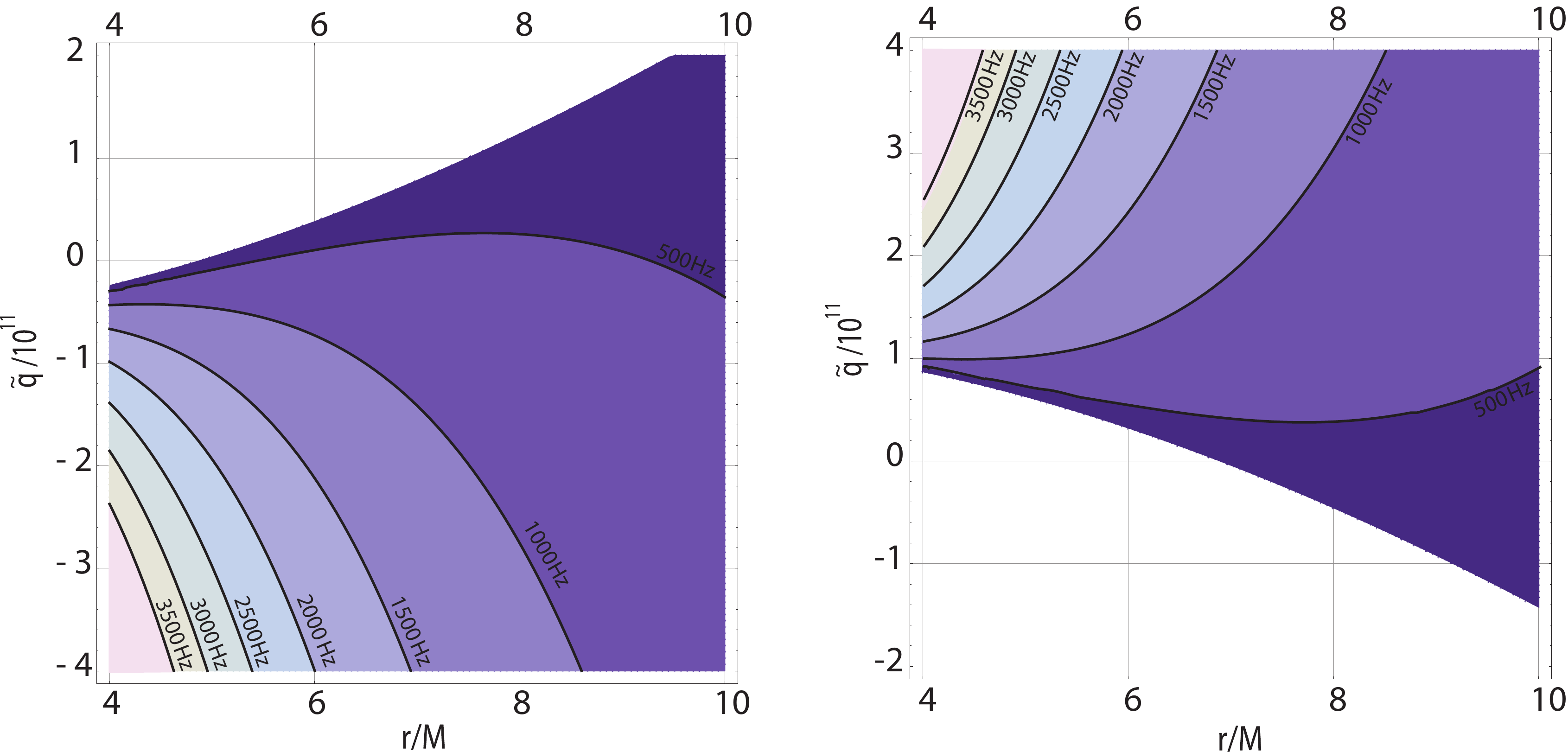}
\includegraphics[width=1.0\hsize, angle=0]{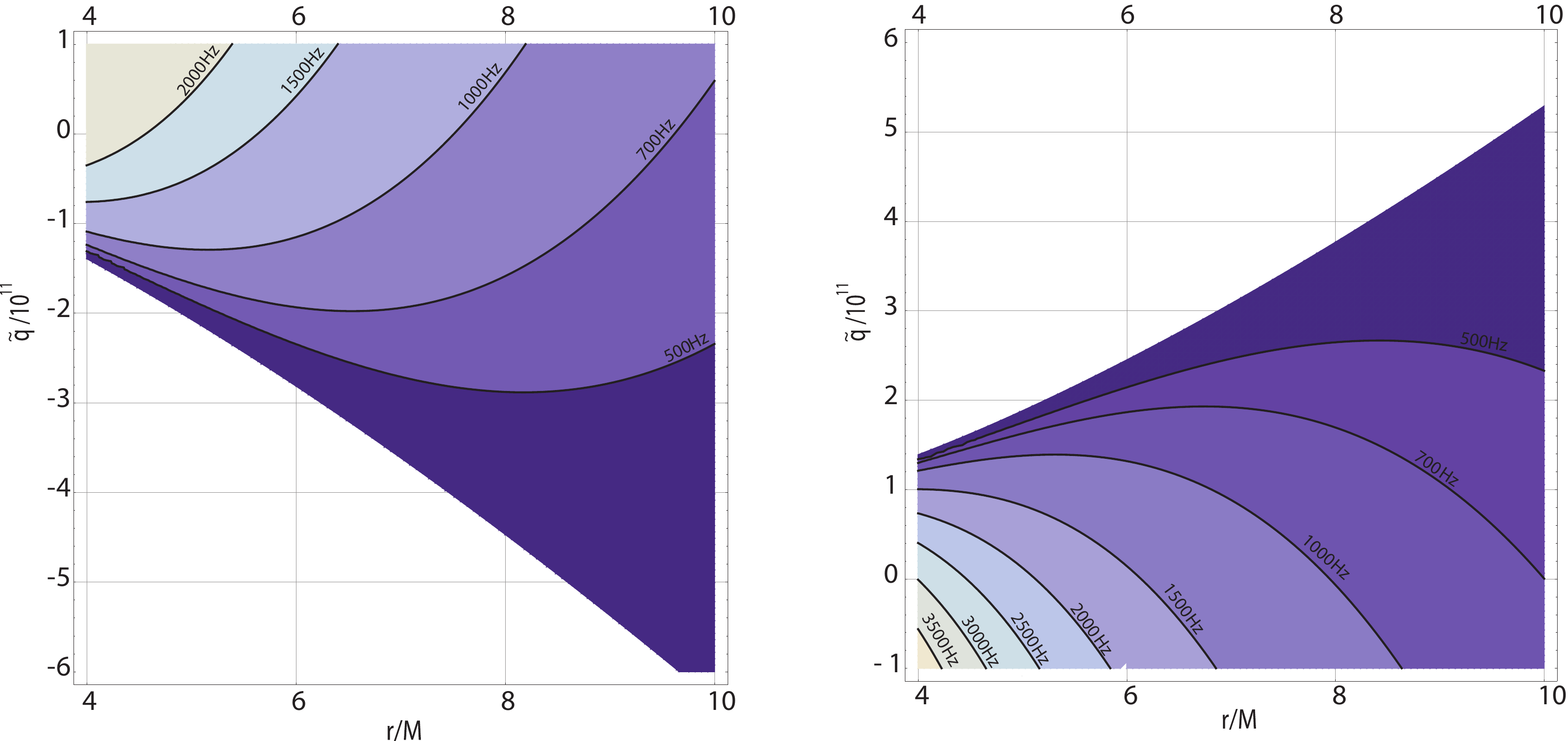}
\noindent
\caption[t]{
{\bf Top left:} Contour plot of the corotating radial epicyclic frequency $\nu_{r\pm}=\omega_{r\pm} / 2\pi$ as a function of the specific charge $\mathrm{\tilde q}$ and the radial coordinate. {\bf Bottom left:} Contour plot of the vertical epicyclic frequency $\nu_{\theta}=\omega_{\theta} / 2\pi$ as a function of the specific charge $\mathrm{\tilde q}$ and the radial coordinate.
{\bf Right panels:} Same as the left panels, but for counter-rotating orbits.}
\label{figure:epi}
\end{figure}

\begin{figure}[t!]
\includegraphics[width=1.0\hsize, angle=0]{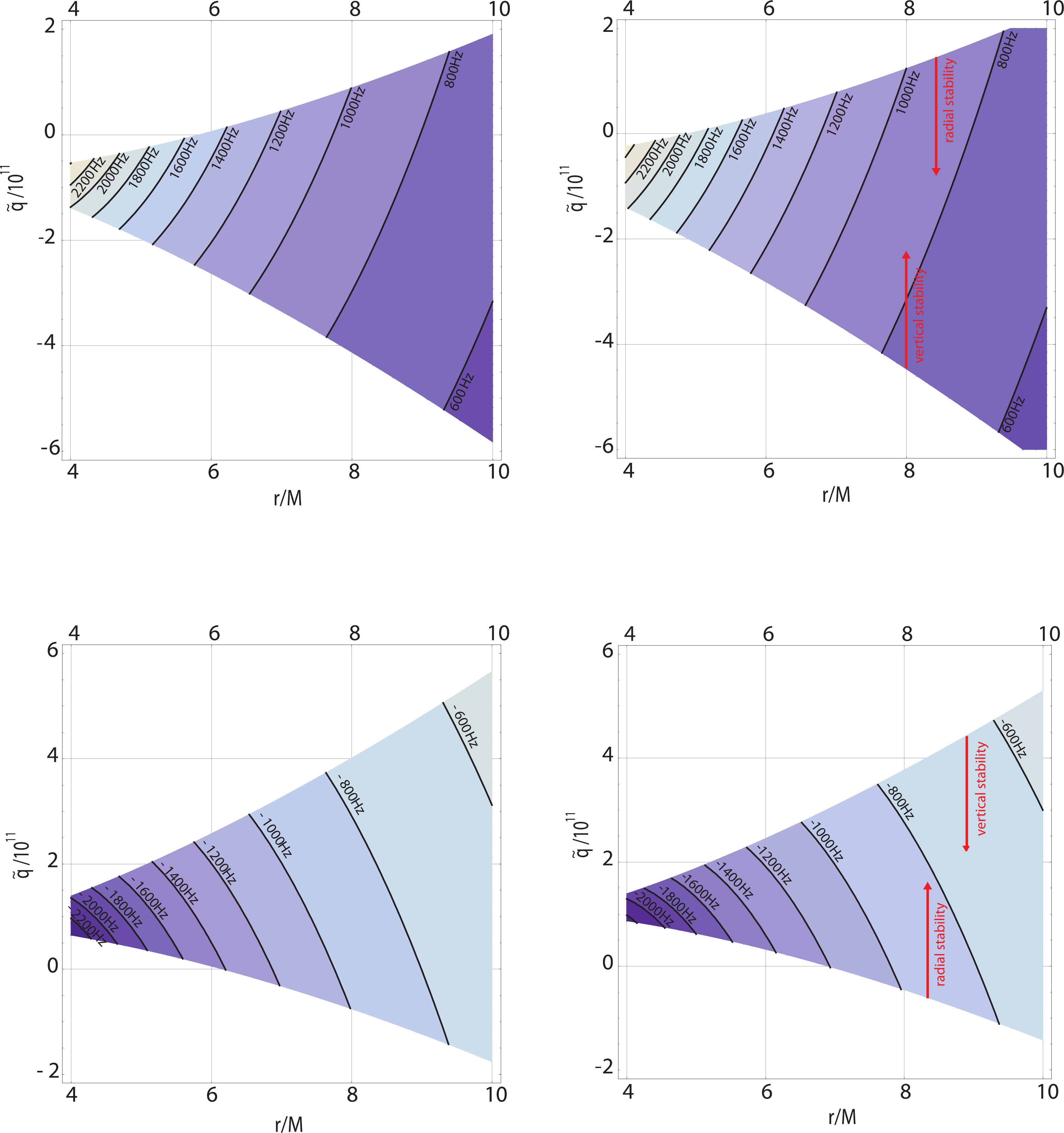}
\noindent
\caption[t]{The region of stable circular orbits filled up by the contour plot of the orbital frequency $\nu=\Omega/2\pi$. Constructed for a test neutron star with  $M=1.5 M_{\odot}$ and $\mu=1.06\,\mathrm{x}\,10^{-4}\,m^{2}$. {\bf Top left:}  Corotating orbits for spin $a=0.05$.  {\bf Top right:} Corotating orbits for spin $a=0.3$.  {\bf Bottom left:} Counter-rotating orbits for spin $a=0.05$.  {\bf Bottom right:} Counter-rotating orbits for spin $a=0.3$.}
\label{figure:stable}
\end{figure}

The existence of epicyclic frequency corresponds to the condition of the stability
relative to the appropriately oriented perturbations. Therefore, vanishing of the radial $\omega_{r}$ or vertical $\omega_{\theta}$ epicyclic frequencies determines the location of the marginally stable circular orbits (Fig. \ref{figure:epi}, \ref{figure:stable}). We use analytic formulae for the radial and vertical epicyclic frequencies of a charged test particle in the presence of a general electromagnetic field derived and discussed by \cite[Aliev \& Galtsov 1981, Aliev 2007, Bakala et al. 2010 \& 2012]{ali-gal:1981, aliev:2007, bak-etal:2010, bak-etal:2012}.

\twocolumn
\vspace{-2ex}
\bigskip

\section*{Acknowledgments}
This work has been supported by the Czech grants GA\v{C}R 202/09/0772 and GA\v{C}R P209/12/P740. The authors further acknowledge the internal student grants of the Silesian University in Opava (SGS/1/2011, SGS/2/2011) and the project CZ.1.07/2.3.00/20.0071 "Synergy" in the frame of "Education for Competitiveness" Czech Operational Programme.

\end{document}